\documentstyle[floats,twocolumn,aps,prl] {revtex}

\begin{document}
\draft

\title{The role of zero-point effects in catalytic reactions involving
hydrogen} 

\author{Axel Gross and Matthias Scheffler}

\address{Fritz-Haber-Institut der Max-Planck-Gesellschaft, Faradayweg 4-6, 
D-14195 Berlin-Dahlem, Germany}

\maketitle

\begin{abstract}

According to the Heisenberg uncertainty principle of quantum mechanics, 
particles which are localized in space by a bounding potential 
must have a  finite distribution of momenta.
This leads, even in the lowest-possible energy state,
to vibrations, and thus, to the so-called zero-point energy.
For chemically bound hydrogen the zero-point energy
can be quite substantial. 
For example, for a free H$_2$ molecule
it is 0.26~eV, a significant value in the realm of chemistry, 
where often an energy of the order of 0.1 eV/atom (or 2.3 kcal/mol)
decides whether or not a chemical reaction takes place with
an appreciable rate.
Yet, in many theoretical studies the dynamics of chemical reactions
involving hydrogen has been treated classically or quasi-classically, 
assuming that the quantum mechanical nature of H nuclei, 
i.e. the zero-point effects, will not strongly
affect the relevant physical or chemical properties. In this paper
we show that this assumption is not justified. We will demonstrate that
for very basic and fundamental catalytic-reaction steps, namely the 
dissociative adsorption of molecular hydrogen at transition metal surfaces 
and its time-reverse process, the associative desorption, 
zero-point effects can not only quantitatively but even qualitatively 
affect the chemical processes and rates. 
Our calculations (treating electrons as well as H nuclei
quantum-mechanically) establish the importance of 
additional zero-point effects
generated by the H$_2$-surface interaction and how energy of the H-H stretch
vibration is transferred into those and vice versa.

\end{abstract}


Recently density functional theory calculations of the 
important diversity of pathways for chemical reactions
at extended metal surfaces
became possible~\cite{Ham94,Whi94,Ham95,Wil95}. Still, these
calculated high-dimensional {\em ab initio} potential-energy surfaces (PES),
which fully determine the energetics of the nuclei, are too complex 
that their  consequences on the chemical reaction dynamics were obvious.
It is worth to stress at this point that these calculations revealed that 
the high-dimensional nature of the PES can be very important, i.e., a
restriction to just one reaction pathway is often not 
sufficient~\cite{Wil95,Gro95}. 
In order to analyze and understand the catalytic dissociation
it is important and possibly even necessary to combine the information 
about the high-dimensional PES of the chemical reaction with a subsequent
calculation of the dynamics of the nuclei, moving along this PES.
Typically this dynamics is described applying the laws
of classical mechanics, the main motivation of this approach being its
simplicity.
Some studies of dissociative adsorption (and associative desorption)
of H$_2$, which is the topic we concentrate on in this paper, 
had addressed the issue of comparing classical and quantum
dynamics~\cite{Eng93,Chi87,Kay95,Kin95}, but
in these studies only up to four degrees of freedom of the hydrogen molecule
were treated quantum mechanically; obviously, the dynamics of two
nuclei is determined by six coordinates.
In the following we will restrict our discussion to the dynamics of 
dissociative adsorption. However, our findings about the importance
of zero-point effects are equally valid for the associative desorption 
since the processes of dissociative adsorption and associative
desorption are related by the principle of microscopic reversibility.

\vspace{-8.0cm}

\hspace*{-1.0cm} {\large {\tt Jour.~Vac.~Sci.~Techn., May~1997}}

\vspace{7.6cm}

\begin{figure}[tb]
\unitlength1cm
\begin{center}
   \begin{picture}(10,9.0)
      \includegraphics{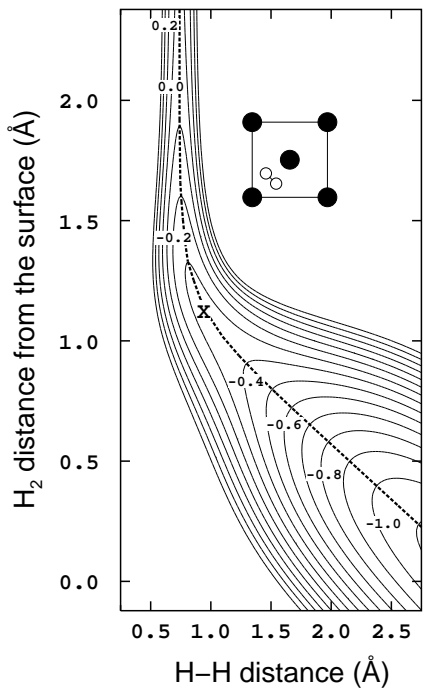}
   \end{picture}

\end{center}
   \caption{Contour plot of the PES along a
            two-dimensional cut through the
            six-dimensional coordinate space of H$_2$/Pd\,(100).
            The inset shows the
            orientation of the molecular axis and the lateral
            H$_2$ center-of-mass coordinates. The coordinates 
            in the figure are the H$_2$ center-of-mass distance 
            from the surface $Z$ and the H-H interatomic distance $d$. The 
            dashed line is the optimum reaction path.
            Energies are in eV per H$_2$ molecule.
            The contour spacing is 0.1~eV.  }

\label{elbow}
\end{figure}

In this paper we will show that a quantum-mechanical,
six-dimensional treatment of the nuclei
exposes  several interesting and important properties.
Such calculations became possible only recently~\cite{Gro95}, and
as a first example we studied the dissociative adsorption
of H$_2$ at the (100) surface of palladium
(for the geometry of the (100) surface see the inset of Fig.~\ref{elbow}).
The employed high-dimensional PES was derived
from density functional theory calculations~\cite{Wil95} which revealed
that it is strongly corrugated and that many activated
and only few non-activated paths towards dissociative adsorption
exist. Figure~\ref{elbow} shows a cut through this PES, which
displays one of the few non-activated pathways towards
dissociated adsorption. In this example the H$_2$ molecule
approaches with an orientation parallel to the surface and
its center of mass at the bridge position (see inset of Fig.~\ref{elbow}).

\begin{figure}[tb]
\unitlength1cm
\begin{center}
   \begin{picture}(10,6.3)
      \includegraphics{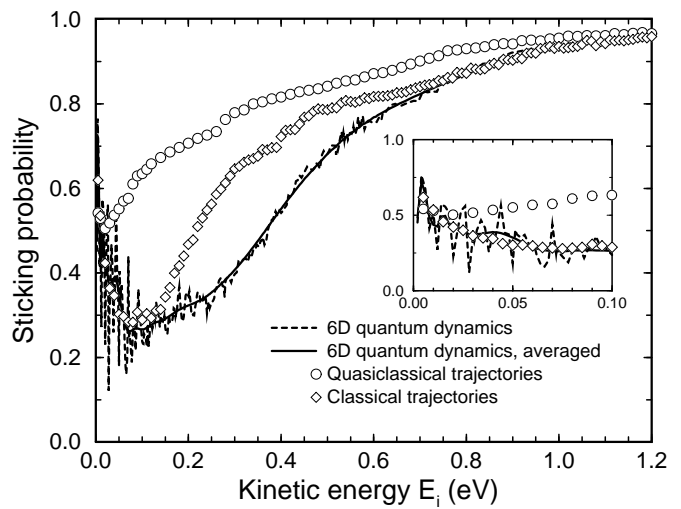}
   \end{picture}

\end{center}
   \caption{Probability for dissociative adsorption
versus kinetic translational energy for a H$_2$ beam under normal
incidence on a clean Pd(100) surface.
The six-dimensional quantum dynamical results (solid and dashed line) 
are for molecules initially in the vibrational 
and rotational ground state. 
For the full line an energy spread typical for beam experiments has
been assumed. Diamonds show results of a classical treatment of the
H nuclei (initially non-vibrating molecules) and circles
correspond to a  quasi-classical treatment, i.e. the molecules 
vibrate initially with an energy of 0.26 eV, which is the
zero-point energy of H$_2$.
The inset shows an enlargement of the results at low-energies.}
\label{stickprob}
\end{figure}

The six-dimensional quantum dynamical calculations show that the
probability for dissociative adsorption of H$_2$ at the clean Pd(100)
is very high  at low kinetic translational energies of the impinging
molecules (up to 70~\%), and that it decreases substantially when the
kinetic translational energy is
increased (to 25 \% at $E_{i} = 0.1$~eV). This result is displayed as the
full curve in Fig.~\ref{stickprob}; it is in fact in good agreement
with experimental findings~\cite{Ren89}. We note that thermal
energies are in the just noted range, i.e. typically $E_{i} \le 0.1$~eV.
The analysis of these results revealed that the decrease of the
sticking probability with increasing kinetic translational
energy is not due to a 
precursor-mediated mechanism, as was commonly believed, but due to
dynamical steering~\cite{Gro95}:
If the impinging molecule is slow the attractive forces of the PES
steer it towards a favorable (non-activated) pathway, even if the initial
conditions were pointing towards an unfavorable direction.
However, if the molecule is fast, its inertia hinders
its reorientation. Such a  molecule will
then probably experience an energy barrier  and thus will be reflected.
Of course, a very fast molecule will simply overrun 
energy barriers.
Another result of the calculations was that the dissociation should
be hindered if low energy molecules are rotationally excited, 
the effect also being dependent on the molecular orientation,
and this prediction has been indeed confirmed by recent
experiments~\cite{Beu95,Wet96}.

In this paper we will discuss the importance of quantum effects
in dissociation and adsorption.
At first we note that calculated H$_2$ sticking probability
exhibits a strong oscillatory structure as a function of the incident
energy (see the dashed line in Fig.~\ref{stickprob}).
These oscillations are a direct consequence of the quantum nature 
of H$_2$ scattering. In a simplified description, i.e., neglecting
rotational and vibrational degrees of freedom of H$_2$, elastic scattering
at a periodic surface gives rise to reflected beams, travelling
in directions  (${\bf k}_{\parallel} + {\bf g},
- \sqrt{ k_{z}^{2} - 2{\bf k}_{\parallel} \cdot {\bf g}
- {\bf g}^{2} }$).
Here $({\bf k}_{\parallel}, k_{z})$ is the wave vector of the 
incident H$_2$ beam and {\bf g} is a two-dimensional
reciprocal lattice vector of the surface. The condition
for emerging beams is that the argument under the square root
is positive. Just before a new beam can emerge, 
it is already built up though it remains confined to the surface.
Thus, this beam can not (yet) be observed directly, but, as it is
coherent with the other beams, these others will exhibit sharp
resonance structures.
In addition, oscillations can also be caused by selective
adsorption resonances \cite{Dar90}.
These structures are known in He and H$_2$ scattering since the 
1930s~\cite{stern} and in electron scattering~\cite{mcrae,pendry},
however, for the particular system H$_2$/Pd(100) they have not been
observed yet.
It is obvious that an experimental verification requires special care
as the strengths and energies of the oscillatory structures depend 
sensitively on the initial conditions, and on the quality of the surface.
For example, surface imperfections (e.g. adatoms and steps) will 
destroy the mentioned coherence and thus suppress the oscillations.

In previous experiments~\cite{Ren89} the molecular beam was 
not strictly mono-energetic. By taking the  experimental
energy spread into account,  $\Delta E / E_i = 0.2$ \cite{Ren89},
the oscillations of the quantum-mechanical study are smoothed out
and the dashed line in Fig.~\ref{stickprob} is turned into
the solid one.

We have now complemented the six-dimensional quantum dynamical study
by classical molecular dynamics simulations  using {\em exactly
the same} PES.
This allows for the first time a thorough comparison of classical
and quantum dynamics of H$_2$ scattering, dissociation, and adsorption.

A common approach  in classical calculations aiming at taking some
quantum effects of the H$_2$ molecule into account is to give the
hydrogen molecules initially a vibrational energy which equals the
H$_2$ zero-point energy~\cite{Eng93,Chi87,Bow89,Mil89}.
This is called a quasi-classical approach.
The open circles in Fig.~\ref{stickprob} display the corresponding results.
It is quite obvious that they are  quantitatively and even
qualitatively at variance with the correct, i.e. quantum-dynamical, result
(dashed curve in Fig.~\ref{stickprob}).
Firstly, the quasi-classical calculations do not show any oscillatory
structure.
Secondly, they do not exhibit any pronounced initial decrease with 
increasing kinetic translational energy, i.e. steering is absent
in the quasi-classical calculations.
Only at very high energies, where quantum effects play a minor role,
quantum and quasi-classical trajectory calculations get into agreement.

The diamonds in Fig.~\ref{stickprob} correspond to a strict classical 
treatment where the H$_2$ molecules have initially no vibrational energy.
It is interesting to note that these calculations
agree much better with the 
quantum-mechanical results than the quasi-classical one.
Again, the quantum oscillations are obviously absent, but
now the steering effect is in operation.
The absence of the oscillatory structure in the classical 
and quasi-classical calculations is caused by the fact that the
coherence and interference of molecular beams is not present in 
classical mechanics.

We will now show that further quantum effects, namely the
generation of additional zero-point energies, lead to the qualitiative 
difference between the quasi-classical and the averaged quantum dynamical
calculations. The interaction of the H$_2$ molecule with the
surface gives rise to these additional zero-point energies (due to
hindered rotations and hindered surface diffusion) and this energy,
obviously, needs to be taken from that of the incident beam.
For the ease of the following discussion we will assume
that the approaching molecule travels along the optimum reaction pathway
(full line in Fig.~\ref{elbow}).
Let us at first describe the correct, i.e. quantum-mechanical, results.
When the molecule gets close to the surface, the energy of its 
stretch vibrations is lowered,
because when the  H-surface bond starts to build up, the
H-H bond strength will at the same time soften.
Since the vibrational frequency is high,
the change of the frequency during one oscillation
is small compared to the frequency, i.e. 
\begin{equation}\label{adiacond}
t_{\rm vib} \ \cdot \ \frac{d \omega_{\rm vib}}{dt} \ \ll \ \omega_{\rm vib}.
\end{equation} 
This suggests that the H$_2$ vibration follows the
molecular center of mass motion adiabatically, which is indeed
confirmed within a very good approximation by the  quantum 
mechanical \cite{Gro96b} as well as by the classical calculations.
Interestingly, the change in zero-point energy is
not transferred to translational motion.
Thus, in contrast to common believe, it cannot be utilized to overcome energy
barriers. The reason is that at the point where
the H-H stretch-frequency is lowered, i.e. where the 
H-H bond is already weakened, also a noticeable H-surface
bond has already started to develop, as was noted already above.
In other words, close to the surface the PES is strongly
corrugated. This leads to the building up of zero-point energies in
the degrees of freedom of the hydrogen molecule 
perpendicular to the optimum reaction path. 
In table~\ref{zeros} all these calculated  zero-point energies are
listed at the point where their sum has its maximum.
This geometry is indicated in Fig.~\ref{elbow} by the cross.
Although the  zero-point energy of the stretch vibrations
is significantly lowered,
this lowering is over-compensated by the zero-point energies in the
other modes so that the sum of all zero-point energies, 0.35~eV per
H-H pair is even larger than the H$_2$ gas-phase value of 0.26~eV.
That the sum of all
zero-points energies in the case of hydrogen dissociative adsorption
is approximately equal or even higher than in the gas-phase has also been
found in total-energy calculations for H$_2$/Al(110)~\cite{Ham92}
and H$_2$/Si(100)~\cite{Kra94}, where values of 0.25~eV/molecule and  
0.29~eV/molecule have been determined at the top of the barrier to
dissociative adsorption.

Obviously, in classical or quasi-classical calculations
the concept of zero-point vibrations does not exist. Therefore in the 
quasi-classical calculations the vibrational energy can be almost 
fully transferred to the translational energy which enables the molecule
to traverse energy barriers with a height larger than its
initial kinetic translational energy. This is actually the reason why
the steering effect is absent in the quasi-classical results
(open dots in Fig.~\ref{stickprob}): These vibrating molecules
do not experience a strong energy barrier because the vibrational
energy is efficiently transferred to translational energy.

\begin{table}
\begin{center}
\begin{tabular}{|c|c|}
mode & zero-point energy (eV)\\
\hline
H-H vibration & 0.126 \\
polar rotation & 0.084 \\
azimuthal rotation & 0.033 \\
translation perpendicular to bridge & 0.051 \\
translation parallel to bridge & 0.056 \\
\hline
sum & 0.350 \\
\end{tabular}
\caption{Zero-point energies of the H$_2$ molecule at the
point where the sum of all zero-point energies has its maximum
(marked by X in Fig. \protect{\ref{elbow}}). 
The H$_2$ center-of mass is above the bridge position between two
adjacent Pd atoms (see inset of Fig. \protect{\ref{elbow}}). 
The energies are given per H$_2$ molecule. 
}
\label{zeros}
\end{center}
\end{table}

The analysis of the zero-point energies along the optimum
reaction pathway (see Fig.~\ref{elbow})
teaches us that the sum of all zero-point energies is roughly constant
during the dissociation.
This approximate ``constant of motion'' can be simulated in classical
trajectory calculations by ignoring the zero-point energies
all along the way. As mentioned above, this strictly
{\em classical treatment} (see the diamonds in
Fig.~\ref{stickprob}) gives results in remarkable agreement with the 
averaged quantum dynamical results at low and high energies. The 
disagreement in the medium energy regime could perhaps be due to the fact
that the sum of the zero-point energies is not really constant
but becomes larger than the gas-phase zero-point energy
so that the quantum mechanical sticking probability is lower
due to the effectively higher barriers.

Our results clearly show that for kinetic translational energies
between  0.15 and 0.6 eV the probability for dissociative
adsorption is reduced by quantum effects by 50~\% compared to the
quasi-classical result. Thus, in contrast to the common belief,
quantum effects in the dynamics of H$_2$ chemical reactions at surfaces
are clearly not negligible.

An influence of the zero-point energies in traversing barrier regions
has also been found in the comparison of three-dimensional quantum and
classical calculations of the dissociative adsorption of
H$_2$/Cu(111)~\cite{Kin95}. 
However, since these calculations were restricted to low dimensions,
the difference between quantum and classical results was much less
pronounced because the sum of the zero-point energies was smaller.
From the above discussion it is clear that a proper
theoretical description of the role of zero-point effects requires a
high-dimensional quantum-mechanical treatment of the nuclear motion.

The problem of a proper treatment of zero-point 
energies in quasiclassical trajectory calculations is 
well-known, especially in the gas-phase community \cite{Bow89,Mil89}. 
One possible way to deal with this problem is the reduced 
dimensionality treatment in the vibrationally adiabatic approximation
(for a overview see Ref.~\cite{Cla86}). In this approach a small 
number of degrees of freedom is treated dynamically while the 
remaining degrees of freedom are taken into account by adding the sum of 
their zero-point energies to the potential along the reaction path.
Another more elaborate approach is to constrain the energy in each
vibrational mode to be greater than its zero-point value \cite{Bow89,Mil89}.

In our purely classical approach we ignore zero-point energies
all along the reaction path. But this approach is actually
in the spirit of the vibrationally adiabatic approximation. 
It effectively takes the zero-point energies into account through 
a shift of the potential along the reaction path corresponding to 
the sum of all zero-point energies. This shift, however, is constant
along the reaction path. Moreover, we still keep the full
dimensionality of the problem by explicitly treating all degrees of 
freedom dynamically. This is indeed essential since for example the
steering effect is  absent in a low-dimensional treatment 
of the H$_2$/Pd(100) system \cite{Dar92} (see also the discussion
below).

\begin{figure}[t]
\unitlength1cm
\begin{center}
   \begin{picture}(10,6.5)
      \includegraphics{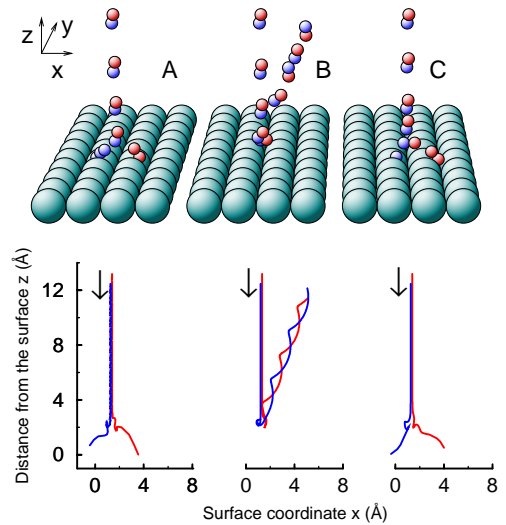}
   \end{picture}

\end{center}
   \caption{Classical trajectories of a hydrogen molecule impinging on a
 Pd(100) surface. The initial conditions are chosen in such a way that the
trajectories are restricted to the $xz$-plane.
Upper panel: snapshots of the trajectories; lower panel: trajectories
in the $xz$-plane. Trajectory A: quasi-classical trajectory with an initial
kinetic translational energy of 0.12~eV. Trajectory B: same initial conditions
as in
trajectory A, except that the molecule is initially non-vibrating. 
Trajectory C: same initial conditions as in trajectory B, except that the 
molecule has a lower kinetic translational energy of 0.01 eV. }

\label{traj3}
\end{figure}

\begin{figure}[t]
\unitlength1cm
\begin{center}
   \begin{picture}(10,7.0)
      \includegraphics{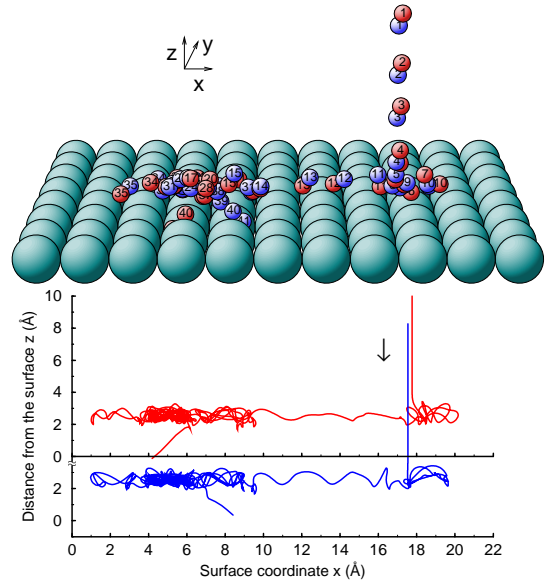}
   \end{picture}

\end{center}
   \caption{Classical trajectory of a hydrogen molecule impinging on a
 Pd(100) surface. The initial kinetic energy is 0.01~eV,
 the initial conditions are only slightly different from those
 of Fig.~\protect{\ref{traj3}}C (see text).
 Upper panel: snapshots of the trajectories which are taken approximately
 every 150~fs and consecutively numbered.
 Lower panel: trajectories in the $xz$-plane; for the sake of clarity
 the trajectories are shifted in the $z$-direction with respect to
 each other.
}
\label{traj1}
\end{figure}

In order to illustrate some of the dynamical effects discussed above 
we have plotted in Figs.~\ref{traj3} and \ref{traj1} some 
typical trajectories of the classical molecular dynamics calculations. 
The upper panel
shows snapshots of trajectories of the impinging H$_2$ molecules
while the lower panel displays the whole trajectories.
The initial conditions are chosen in such a way that the trajectories are
restricted to a $xz$-plane. Trajectories A and B 
in Fig.~\ref{traj3} illustrate the
difference between quasi-classical and classical calculations.
They have identical initial conditions with a kinetic translational
energy of 0.12~eV
except for the fact that in case A the molecule initially
vibrates with an energy corresponding to the zero-point energy while
in case B the molecule is initially non-vibrating (due to the small
amplitude of 0.12~{\AA} the vibrations are hardly perceivable 
on the scale of the figure). For the chosen initial conditions 
of the impinging molecules there is an energy
barrier hindereing the dissociative adsorption, and
the initial translational energy is not large enough for the 
molecule to overcome this.
Thus,  the molecule is scattered back at the repulsive
potential (trajectory B), whereas if the initial vibrational zero-point
energy is transferred to the translation due to the adiabaticity
of the vibrations (trajectory A), the molecule can directly dissociate.

Finally, trajectory C illustrates the steering effect \cite{Gro95,Kay95}. 
The initial conditions are as in trajectory B, but now the molecule has 
a much lower kinetic translational energy, $E_i = 0.01$~eV. 
The molecule is so slow that the attractive
forces can reorient the molecule so that it can follow a non-activated
path towards dissociative adsorption. We like to emphasize that steering 
works in all geometry parameters of the nuclei and becomes more important 
when the configuration space is high-dimensional -- in a one-dimensional 
description (just restricting the reaction to ``the optimum'' reaction 
pathway) or a two-dimensional description (restricting the dynamics to 
only one elbow \cite{Dar92}), the steering effect is absent. 

Trajectory~C in Fig.~\ref{traj3}  represents a {\em classical}
illustration of steering. We note that this picture also applies 
when the quantum nature of H$_2$ molecules is considered, and we now 
briefly address the quantum mechanical picture. Let us assume that the 
incident H$_2$ molecule is in the rotational ground state, 
$j_{i} = 0$ (the description for rotationally excited states 
is analogous). The $j_i =0$ state refers to an isotropic distribution 
of molecular orientation. Thus, the molecule is 
not rotating, but when the orientations of different
molecules in the  beam were measured, one would find all orientations 
with the same  probability.
In other words, the probability that a molecule impinges with the
favorable orientation noted in Fig. 1 is very low.
For molecules with low kinetic energy the steering effect will align all
molecules, and trajectory~C shows one example, which acts on the molecular
polar coordinate $\theta$. Equivalent processes take place 
in the azimuthal $\phi$ and the $(X, Y)$ coordinates. Also with
respect to the $(X, Y)$  coordinates of the H$_2$ molecule we note their
meaning in (basic) quantum mechanics: A plane wave of hydrogen molecules
representing a molecular beam impinging under normal incidence on a surface
has no $(X,Y)$ dependence in the gas-phase. This means that the probability 
to find a molecule at any given point point in a $(X, Y)$ plane far away
from the surface is the same. This brief reminder to the concepts of 
quantum mechanics highlights also
the big  advantage of quantum dynamics in comparison to molecular dynamics.
A quantum dynamical approach includes the ensemble of all H$_2$ orientations
($\theta, \phi$) and all $(X, Y)$ points of impact automatically. In a
molecular dynamics
calculations the same ensemble has to be built up step by step, i.e. by
performing thousands of molecular dynamics runs, namely one run for
each choice of $(X, Y, \theta, \phi)$.

The trajectory in Fig.~\ref{traj1} finally indicates 
a process which in a quantum-mechanical system will give rise to
an oscillatory structure (see Fig.~\ref{stickprob}).
The initial conditions are as in the trajectory 
of Fig.~\ref{traj3}C, only the impact point in the surface unit cell 
and the inital polar orientation are slightly changed. 
The snapshots of the trajectory in the upper panel are
taken approximately every 150 fs. They are consecutively numbered
so that the dynamics can be roughly followed.
Due to the strong anisotropy and the corrugation 
of the potential the translational energy perpendicular to the surface
is efficiently transferred to rotational and kinetic 
energy parallel to the surface. Hence the molecule is temporarily
trapped; in this case the molecule spends
more than 5~ps in front of the surface {\em before it is dissociated}.
Note that this trapping does not occur
because of energy transfer to the surface but because of the energy
redistribution among the H$_2$ degrees of freedom. 
Quantum mechanically the trapped molecules correspond to
metastable states at the surface which only exist for certain
discrete energies; in addition, only discrete 
values for the energy and momentum transfer are allowed.
Thus only for particular initial energies the dynamical trapping into 
these so-called selective adsorption resonances is 
possible\cite{Dar90,stern}.
As mentioned in the discussion of Fig.~\ref{stickprob},
these resonances in addition to the opening up of scattering channels
lead to the sharp peaks in the quantum-mechanical sticking probability.

In conclusion, we have shown that there are large
differences between classical and quantum dynamics of hydrogen.
The classical results do not show any oscillatory structure
which in the quantum dynamics is due to the opening up of new 
scattering channels and resonances. Even more remarkable,
zero-point effects can cause substantial deviations between classical 
and {\em averaged} quantum dynamical calculations.
A corrugated and anisotropic potential energy surface
leads to the building up of zero-point energies 
which effectively increase the minimum potential in the quantum  
calculations. This can change the dynamics of chemical reactions
not only quantitatively, but also qualitatively, and thus strongly
affects the rate constants of catalytic reactions involving
hydrogen. These zero-point effects will also be relevant for
other studies of hydrogen dynamics in corrugated potentials,
like for example in hydrogen diffusion on surfaces or in the bulk.
If, however, the sum of the zero-point energies is approximately constant 
during the traversing of regions with energy barriers,
which for hydrogen dissociation on surfaces
seems to be the case for a wide class of systems,
the discrepancies between classical and averaged
quantum dynamics can be diminished by neglecting zero-point energies in the
initial conditions.

\end{document}